\documentclass[conference]{IEEEtran}
\usepackage{cite}
\usepackage{float}  
\usepackage{amsmath,amssymb,amsfonts}
\usepackage{algorithmic}
\usepackage{graphicx}
\graphicspath{{pictures/}}
\usepackage{textcomp}
\usepackage{xcolor}
\def\BibTeX{{\rm B\kern-.05em{\sc i\kern-.025em b}\kern-.08em
    T\kern-.1667em\lower.7ex\hbox{E}\kern-.125emX}}
\begin{document}

\title{Analyzing the Effects of CI/CD on Open Source Repositories in GitHub and GitLab}

\author{\IEEEauthorblockN{Jeffrey Fairbanks}
\IEEEauthorblockA{\textit{Department of Computer Science} \\
\textit{Boise State University}\\
Boise, Idaho, USA \\
jeffreyfairbanks@u.boisestate.edu}
\and
\IEEEauthorblockN{Akshharaa Tharigonda}
\IEEEauthorblockA{\textit{Department of Computer Science} \\
\textit{Boise State University}\\
Boise, Idaho, USA \\
akshharaatharigo@u.boisestate.edu }

\and
\IEEEauthorblockN{Nasir U. Eisty}
\IEEEauthorblockA{\textit{Department of Computer Science} \\
\textit{Boise State University}\\
Boise, Idaho, USA \\
nasireisty@boisestate.edu }
}
\maketitle

\begin{abstract}


Numerous articles emphasize the benefits of implementing Continuous Integration and Delivery (CI/CD) pipelines in software development. These pipelines are expected to improve a project's reputation and decrease the number of commits and issues in the repository. Although CI/CD adoption may be slow initially, it is believed to accelerate service delivery and deployment in the long run. This study aims to investigate the impact of CI/CD on commit velocity and issue counts in two open-source repositories, GitLab and GitHub. By analyzing more than 12,000 repositories and recording every commit and issue, it was discovered that CI/CD enhances commit velocity by 141.19\% but also increases the number of issues by 321.21\%.

\end{abstract}


\section{Introduction}

It's rare for resolving code issues to be a simple task. Unforeseen problems and unanticipated test cases often arise during the project's development. This causes the project to shift its original intentions and require different inputs and outputs. Keeping track of these changes can be challenging, mainly when working with multiple developers, and it may increase the resources needed for the project.

Developing a project with a team of developers requires effective communication and a simple method for implementing new code with new test cases. Unfortunately, in many cases, testing code frequently and in small increments is disregarded until after a significant amount of code has been written. Unit testing has gained popularity in recent years and allows for more efficient testing as code is developed. However, it is often employed only as the code section conducting unit testing is nearing completion.

As testing (both unit and end-to-end) is an integral part of the development life cycle, a newer solution has come to test its validity. The software engineering community widely advocates continuous integration and Continuous Delivery (CI/CD). There are two parts to CI/CD.

CI (Continuous Integration) is a practice that helps many developers integrate code changes into a single project more quickly and easily. Developers use an automated set of tests to validate the code before it's integrated or merged into a project's main branch or a repository that is shared with other developers on the team. This allows development teams to build quickly and test their code continuously, which helps them to avoid problems later on.

On the other hand, CD (Continuous Delivery) is a great way to automate the code changes to the process. It is a better way to deliver changes to development or production environments than traditional methods. Continuous Development can also push rapid and small changes to various environments, allowing for quick rollbacks or changes and avoiding major differences or exceptions. When using and pushing small changes, deploying and testing code that could be troublesome becomes substantially more straightforward. In addition, this method allows for risk mitigation when pushing to a critical production system.

CI/CD is essential in a development pipeline as it enables a faster development cycle, facilitates the identification of imminent issues with the code, and promotes efficient and minimal code changes. By gathering and presenting essential metrics on the benefits of utilizing CI/CD in a development pipeline, developers can enhance their productivity and deliver exceptional products to users. Consequently, our research endeavors to address the following inquiries: 

\textbf{RQ1: Does integrating CI/CD enhance commit velocity?}

\textbf{RQ2: Does integrating CI/CD impact the number of issues reported for the project?}

\textbf{RQ3: Are there any significant variations in CI/CD issues between GitLab and GitHub?}

We designed these inquiries to evaluate the effectiveness of utilizing CI/CD in projects involving developers' teams. Although many articles assert that CI/CD improves commit velocity and streamlines testing and deployment pipelines, limited formal research substantiates these claims~\cite{b25, b26, b27}. Moreover, utilizing CI/CD requires implementing multiple tools and processes that must work together harmoniously. Correctly learning and implementing these tools can be time-consuming and challenging, leading many developers to question their value \cite{b2}.

The primary focus of the third research question is to examine whether there are significant disparities in the utilization of CI/CD between GitLab and GitHub. This metric holds paramount importance as one platform may facilitate commit velocity while the other may impede the advantages of utilizing CI/CD. Similarly, the drawbacks associated with CI/CD, such as the time-consuming setup process, may vary between the two open-source repository platforms. This study aims to investigate these variations and determine if any noteworthy differences exist that could pose challenges in implementing CI/CD.

To answer these inquiries, we collected data from over 12,000 repositories, including those with and without CI/CD. We mined data from open-source repositories on GitLab and GitHub, gathering information on projects with multiple developers and CI/CD workflows using their respective APIs. We analyzed the data and answered our research questions using custom functions found in Google Sheets. Additionally, we utilized the programming language Python and the Google Colab platform to analysis and process the data.

\section{Related work}
Today's development climate is marked by research on CI/CD that focuses on transitioning to a CI/CD pipeline, pinpointing pain points in its usage, or comparing different platforms that use CI/CD (e.g., comparing the use of GitLab versus GitHub) \cite{b28, b29, b30}.

Savor et al.~\cite{b5} identified four critical elements of continuous deployment: small software updates, automatic releases, developer responsibility, and fully automatic deployment. Tools aid in managing these responsibilities by offering insights into code changes, automating repetitive and error-prone tasks, and logging every process action. Their findings suggest that CI/CD should be a continuous and automated process that continuously tests and deploys code to production. Based on this understanding, our research aims to ensure that CI/CD, as used in real-world applications, actually increases development velocity as intended.

In their study, Rangnau et al.~\cite{b3} explore the integration of security tools within CI/CD pipelines. While they assume that CI/CD is widely used today, no formal research supports this claim. Their work contributes to our study by highlighting the shift towards a SaaS (Software as a Service) development model, where multiple users share instances running on cloud infrastructure. This new development environment allows software practitioners to continuously enhance product quality through frequent updates, leading to increased use of CI/CD. Our research examines the validity of this claim by analyzing a real-world and up-to-date dataset.

Nogueira et al.~\cite{b4} demonstrate how Apache Kafka pipelines support integrating and normalizing event logs from multiple sources into data streams that feed process mining algorithms in real-time. They apply this to the complex CI/CD pipeline of a major European e-commerce company, highlighting how these techniques enhance the monitoring and observability of development processes. Their study focuses on a single CI/CD pipeline, whereas our research examines multiple pipelines to investigate whether velocity is improved overall and if development issues are reduced.

\section{Methodology}
This section discusses the data collection and analysis methodology employed in this research. The study involved mining two primary open-source repositories to obtain information on the usage of CI/CD. Specifically, we investigated the GitHub and GitLab platforms to identify any variations in velocity and CI/CD utilization between them. Despite their apparent similarities, these platforms exhibit distinct differences~\cite{b16} that we aimed to uncover. Therefore, in the following discussion, we will delve into the data collection and analysis process for each platform.

\begin{figure}[htbp!]
    \centering
    \caption{Overview of the Research Methodology}
    \label{fig:overview}
    \includegraphics[width=9cm]{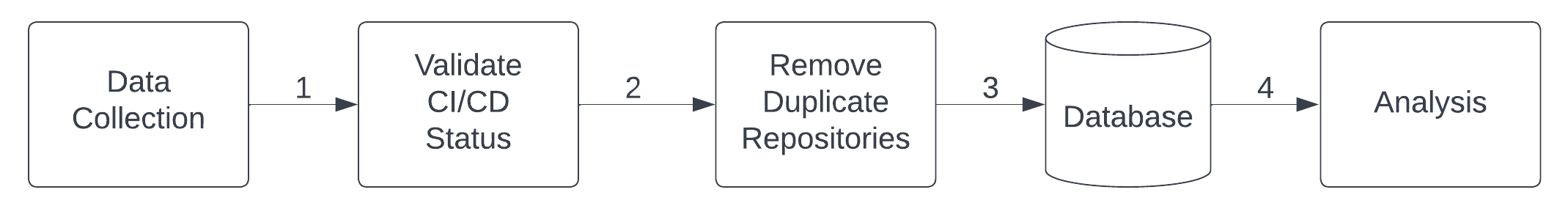}
\end{figure}
 
Fig.~\ref{fig:overview} depicts the general process we followed for each of the open-source repositories, GitHub and GitLab. The first step involved collecting data and verifying whether each repository had CI/CD capabilities. Once this was established, we cleaned the data and eliminated any duplicate entries, if present. Subsequently, we stored the cleaned data in a database and performed further analysis to address the research questions.

\subsection{Collecting Data from GitHub }

\begin{figure}[htbp!]
    \centering
    \caption{GitHub Data Collection Method 1}
    \label{fig:githubCollection1}
    \includegraphics[width=9cm]{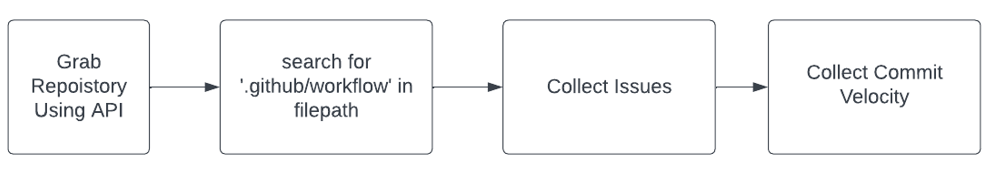}
\end{figure}

GitHub is an open-source repository for developers to store, collaborate, and manage their code in a centralized location~\cite{b17}. To collect data from GitHub, we primarily utilized the API provided by the platform. We adopted several different approaches while using the API. Initially, we collected information on repositories that implemented CI/CD and then further mined this subset of repositories to gather data on pull requests, issues, and related metadata. To achieve this, we utilized the GitHub API to search the path of each publicly available repository for the file path: ``.github/workflow"~\cite{b18}. Fig.~\ref{fig:githubCollection1} shows an overview of this data collection method. 

This search was conducted to identify projects that contained CI/CD since this file is required to implement CI/CD within a repository. This method was effective until we reached the limit on the number of results. As per the documentation, only 1000 results can be returned for each search~\cite{b19}. Thus, we needed another approach to collect the required CI/CD projects after reaching the limit of 1000 repositories.

\begin{figure}[htbp!]
    \centering
    \caption{GitHub Data Collection Method 2}
    \label{fig:githubCollection2}
    \includegraphics[width=9cm]{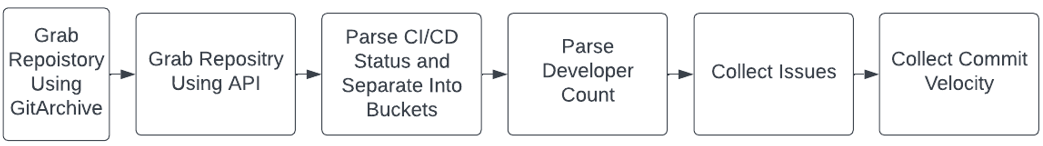}
\end{figure}

To collect data from GitHub, we adopted another approach of scraping a list of developer name/repository name pairs using GitArchive. After compiling a substantial list, we utilized the GitLab API to check each repository for CI/CD. Next, we parsed each repository into a "using CI/CD" or a "not using CI/CD" bucket. To prevent personal or school projects from skewing the data, we pruned each bucket to include only repositories with at least two developers. We also ensured that each repository was active in 2022, thus ensuring the data is current and valuable for this research. This guarantees that the results produced by this research project are relevant and applicable to the current time. Fig.~\ref{fig:githubCollection2} shows an overview of this data collection method.

\begin{figure}[htbp!]
    \centering
    \caption{GitHub Data Collection Method 3}
    \label{fig:githubCollection3}
    \includegraphics[width=9cm]{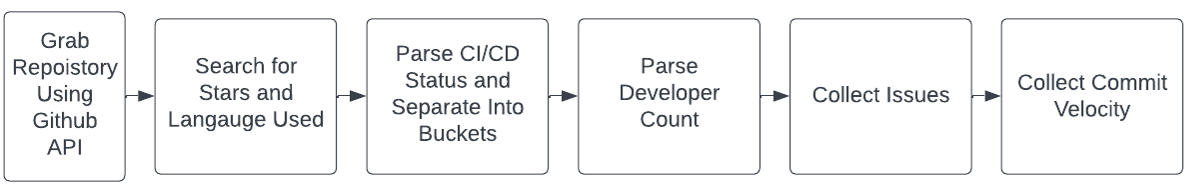}
\end{figure}

We used another search method using the GitHub API to collect the remaining data from GitHub. We formulated the search by filtering repositories by language (e.g., python, java) and sorting them based on the number of stars. This method proved highly effective, as we observed a strong correlation between the number of stars and the implementation of CI/CD pipelines in the project repositories. The more stars a repository had, the more likely it was to have implemented CI/CD. This search method enriched our dataset significantly, boosting our confidence in the quality of the data collected. Fig.~\ref{fig:githubCollection3} provides an overview of this data collection approach.

We conducted a deduplication process on all the collected repositories to ensure that each repository was collected only once and categorized correctly based on whether it implemented CI/CD. Then, we ran a validation function on each repository to ensure it was in the appropriate bucket. This step was important to ensure the accuracy and integrity of the data. After the deduplication process, it was confirmed that each repository was in the correct category and that there were no duplicates. 

\textbf{Issues With GitHub Data Collection.} The primary challenges encountered during the GitHub data collection were related to the hourly rate limit imposed by the API when searching for repositories. This limitation was not ideal and prolonged the data collection process significantly. Another issue was that each API query could only return up to 1000 results, meaning multiple searches had to be conducted for each bucket, "using CI/CD" and "not using CI/CD." This led to considerable time being spent validating and de-duplicating the data.

\subsection{Collecting Data from GitLab }
\begin{figure}[htbp!]
    \centering
    \caption{GitLab Data Collection Method}
    \label{fig:gitlabCollection}
    \includegraphics[width=9cm]{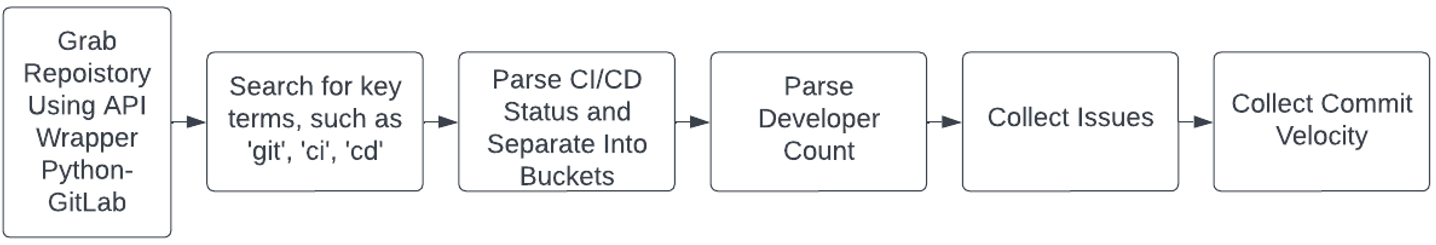}
\end{figure}

The process of collecting data from GitLab differed significantly from collecting data from GitHub. Initially, we attempted to utilize GitLab's own resources for data collection~\cite{b22}. However, we found that it took approximately 5 minutes to collect data from a single repository, which was deemed too time-consuming for the project's timeline. Consequently, we refined our approach by using a python wrapper instead of the API natively~\cite{b23}. Although the wrapper did not support global searching of a specific file, we devised a workaround. We used the python GitLab API wrapper to search for keywords found in the repository name across GitLab. Therefore, we searched for relevant data using keywords such as 'ci', 'cd', 'git', and 'workflow'. Fig.~\ref{fig:gitlabCollection} provides an overview of this data collection approach.

The method we used for data collection was not optimal in returning results. We had to spend a considerable amount of time to ensure that the collected data met the research's standards. We invested a significant amount of time ensuring that the repositories had multiple developers, were categorized correctly based on their use of CI/CD and that only unique entries were collected. This time investment was critical to ensure that we collected high-quality data. Although we attempted other implementations, this approach proved to be the only reliable way to collect the required amount of data at scale within the research's timeline.

\textbf{Issues With GitLab Data Collection.} The process of gathering data from GitLab is significantly more arduous, particularly in terms of searching for relevant information, which consumes a considerable amount of time. In comparison to collecting data from GitHub, obtaining GitLab data takes considerably longer. Despite the need to employ multiple methods to collect GitHub data, it is still roughly three times faster than gathering GitLab data. The disparity in speed is largely due to GitLab's requirement to perform an API search for each issue and timestamp on commits individually, as opposed to GitHub, which can retrieve all necessary information with a single API call. Consequently, the API call-intensive process imposed by GitLab restricts the number of repositories that can be collected in a day and incurs significant waiting time as the API replenishes requests. 

An additional problem encountered when searching for repositories on GitLab is that the API lacks the comprehensive functionality offered by GitHub's API. Compared to GitHub, the GitLab API is less user-friendly and has fewer resources available to facilitate the search for repositories on the platform. In addition, unlike GitHub, the GitLab API does not support the search for file paths, highly-rated projects, language-based projects, or projects that use CI/CD. Consequently, a new approach was necessary to gather data on projects utilizing CI/CD on GitLab. 

While searching for a new data collection method, we discovered that many researchers began by scraping repository URLs and then enriching the data using the GitPy tool to create a temporary local clone of the project. The logs required for the research were extracted from the local clone, and the project was then deleted~\cite{b20}. Another method we found during our search involved crawling GitLab to collect the required data~\cite{b21}. However, neither of these methods was ideal for collecting large amounts of data. Consequently, we decided to develop a new data collection method from scratch, as described above.

\subsection{The Data Set}

We collected data throughout this research, which comprises about 12,000 repositories. For GitHub data, there were at least 3000 repositories in each category; for GitLab data, roughly 1300 repositories were in each category. The collected data enabled us to analyze to answer the research questions. However, in this section, we will demonstrate the data to provide the reader with insight into the collected data. Table~\ref{tab:number_of_repositories} represents the number of repositories with and without CI/CD.

\begin{table}[H]
\centering
\renewcommand{\arraystretch}{1.3}
\caption{Number of Repositories}
\label{tab:number_of_repositories}
\begin{tabular}{|c|c|c|}
\hline
 &   \textbf{CI/CD} &   \textbf{NO CI/CD} \\ \hline
\textbf{GitHub}    &  3223   &  6007  \\ \hline
\textbf{GitLab}    &  1357   &  1356  \\ \hline
\end{tabular}
\end{table}

We collected a few attributes associated with each repository. We cataloged each repository based on its file path in GitHub or GitLab. Additionally, we stored the number of active or retired issues within the lifespan of each repository. We collected the issues to analyze whether CI/CD creates an environment where fewer issues are created. Furthermore, we collected the average time between commits. We calculated the commit velocity of the repository by determining the mean between the time of each commit, and we calculated each commit to obtain an accurate measure. As shown in Table~\ref{tab:data_layout}, we designed the layout of our collected data. We further parsed the data above to enable calculations to take place. We subsequently analyzed the data using various statistical measures, and we describe the results in the next section.

\begin{table}[H]
\centering
\renewcommand{\arraystretch}{1.5}
\caption{Layout of the Data}
\label{tab:data_layout}
\begin{tabular}{|c|c|c|}
\hline
\textbf{Repository Name} & \textbf{Issues} & \textbf{Commit Velocity} \\ \hline
  requilence/integram/  &  64  &  639:10:58  \\ \hline
  bueltge/WordPress-Admin-Style/ &  2  &  494:15:37  \\ \hline
 apache/airflow/ &  863   &  4:09:41  \\ \hline
  openwdl/wdl/  &  65   &  872:34:25  \\ \hline
  microg/android\_packages\_apps\_UnifiedNlp/ &  97  &  580:26:51  \\ \hline
 Froxlor/Froclor/ &  32  &  4:10:46 \\ \hline
   cloudson/gitql/   &  4   &  251:57:46 \\ \hline
  repoze/repoze.workflow/ &  4   &  1679:15:44 \\ \hline
 john30/ebusd/ &  60   &  39:40:22 \\ \hline
   KSP-CKAN/NetKAN/   &  27   &  24:42:51  \\ \hline
  snarfed/bridgey/ &  95  &  31:35:57  \\ \hline
\end{tabular}
\end{table}

\section{Results}

In this section, we will present and discuss the findings of this study. We used the research questions to guide the discussion and discovered several notable outcomes through this study. Finally, we will discuss the points collected between using CI/CD on GitHub and GitLab. 

First, we conducted an analysis of the GitHub collection data to find the mean, median, and standard deviation within the dataset. Initially, we focused on finding the distribution of issues between the projects that utilized CI/CD and those that did not implement the process. After collecting and resourcing the open-source repositories, we conducted a more detailed analysis of each metric within the repository.


\subsection{Repository Issues Analysis}

After analyzing the data on a macro level, we took a much more micro approach to find the small details hiding within the data set. We used this micro approach to mine the repository data, giving us a better look at the distribution of the number of issues found in projects that use CI/CD and those that do not use CI/CD. We analyzed the entire data set and formulated the mean, standard deviation, and median of the number of issues within it as shown in Table~\ref{tab:GitHub_Repository_Issues}.

\begin{table}[H]
\centering
\renewcommand{\arraystretch}{1.3}
\caption{GitHub Repository Issues}
\label{tab:GitHub_Repository_Issues}
\begin{tabular}{|c|c|c|}
\hline
 &   \textbf{CI/CD} &   \textbf{NO CI/CD} \\ \hline
\textbf{Mean}    &  175.38   &  52.33  \\ \hline
\textbf{Median}    &  29   &  25 \\ \hline
\textbf{Standard Deviation}    &  549.97   &  90.55  \\ \hline
\textbf{Count}    &  3223   &  3223 \\ \hline
\end{tabular}
\end{table}



Throughout this study, we also collected and analyzed GitLab data. We demonstrate the results of the distribution between the number of issues for each in regard to CI/CD usage in Table~\ref{tab:GITLAB_REPOSITORY_ISSUES}.

\begin{table}[H]
\centering
\renewcommand{\arraystretch}{1.3}
\caption{GitLab Repository Issues}
\label{tab:GITLAB_REPOSITORY_ISSUES}
\begin{tabular}{|c|c|c|}
\hline
 &   \textbf{CI/CD} &   \textbf{NO CI/CD} \\ \hline
\textbf{Mean}    &  52.04   &  17.56  \\ \hline
\textbf{Median}    &  12  &  3 \\ \hline
\textbf{Standard Deviation}    &  477.56   &  189.81  \\ \hline
\textbf{Count}    &  1357   &  1357 \\ \hline
\end{tabular}
\end{table}

Upon examining the data, it is evident that repositories without CI/CD processes have a lower number of issues. This conclusion is drawn from the lower mean of issues found within the repository, combined with a standard deviation that is significantly lower than that of the repositories containing CI/CD. Therefore, we can conclude that, on average, repositories without CI/CD have fewer issues compared to those with CI/CD in the development process.




\subsection{Repository Commit Velocity}
After analyzing the number of issues in repositories from both GitHub and GitLab, we further investigated the commit velocity between the two platforms. It is worth noting that the mean and median values presented in Tables~\ref{tab:{GitHubRepositoryCommitVelocity}} and \ref{tab:{GitLabRepositoryCommitVelocity}} are calculated in hours. Upon examining the GitLab data, it becomes apparent that the commit velocity shows a similar trend to what we observed on GitHub. On average, the commit velocity is significantly lower when CI/CD is used in the pipeline, compared to when it is not used.

\begin{table}[H]
\centering
\renewcommand{\arraystretch}{1.3}
\caption{ GitHub Repository Commit Velocity}
\label{tab:{GitHubRepositoryCommitVelocity}}
\begin{tabular}{|c|c|c|}
\hline
 &   \textbf{CI/CD} &   \textbf{NO CI/CD} \\ \hline
\textbf{Mean}    &  16.51   &  27.11  \\ \hline
\textbf{Median}    &  18:39:28  &  487:37:01 \\ \hline
\textbf{Standard Deviation}    &  52.05   &  31.92  \\ \hline
\textbf{Count}    &  3223   &  3223 \\ \hline
\end{tabular}
\end{table}


\begin{table}[H]
\centering
\renewcommand{\arraystretch}{1.3}
\caption{ GitLab Repository Commit Velocity}
\label{tab:{GitLabRepositoryCommitVelocity}}
\begin{tabular}{|c|c|c|}
\hline
 &   \textbf{CI/CD} &   \textbf{NO CI/CD} \\ \hline
\textbf{Mean}    &  22.01   &  25.70  \\ \hline
\textbf{Median}    &  32:39:15  &  120:57:14 \\ \hline
\textbf{Standard Deviation}    &  88.57   &  60.21 \\ \hline
\textbf{Count}    &  1356   &  1356 \\ \hline
\end{tabular}
\end{table}

Upon analyzing the commit velocity data from both GitHub and GitLab, we observed that the projects using CI/CD have a faster commit velocity compared to those that do not use CI/CD. This trend is demonstrated in Tables~\ref{tab:GitHubCommitVelocityMetrics} and \ref{tab:GitLabCommitVelocityMetricss}, as expected.

\begin{table}[H]
\centering
\renewcommand{\arraystretch}{1.3}
\caption{GitHub Commit Velocity Metrics}
\label{tab:GitHubCommitVelocityMetrics}
\begin{tabular}{|c|c|c|}
\hline
 &   \textbf{CI/CD} &   \textbf{NO CI/CD} \\ \hline
\textbf{Mean}    &  16.51 Hrs   &  27.11 Hrs\\ \hline
\textbf{Standard Deviation}    &  52.05    &  31.92 \\ \hline
\textbf{Median}    &  18 Hrs   &  487 Hrs  \\ \hline
\end{tabular}
\end{table}

\begin{table}[H]
\centering
\renewcommand{\arraystretch}{1.3}
\caption{GitLab Commit Velocity Metrics}
\label{tab:GitLabCommitVelocityMetricss}
\begin{tabular}{|c|c|c|}
\hline
 &   \textbf{CI/CD} &   \textbf{NO CI/CD} \\ \hline
\textbf{Mean}    &  22.01 Hrs   &  25.7 Hrs\\ \hline
\textbf{Standard Deviation}    &  88.57    &  60.21 \\ \hline
\textbf{Median}    &  32 Hrs   &  120 Hrs  \\ \hline
\end{tabular}
\end{table}

The most significant finding regarding the use of CI/CD and commit velocity is the median time between commits. On average, when using CI/CD in their pipelines, the median commit velocity is significantly lower compared to those that do not use CI/CD. Based on this analysis, we can conclude that implementing CI/CD results in an increase in commit velocity. 

\section{Discussion}

Through the implementation of this research, a clear outcome has been achieved and the research question are ready to be answered. Each of the questions will lead the discussion in this section of the paper. 

\subsection{\textbf{RQ1: Does integrating CI/CD enhance commit velocity?}}

The data clearly shows that implementing CI/CD increases the commit velocity on average, which is the primary goal of CI/CD. On average, the implementation of CI/CD results in a significant improvement in commit velocity of 141.19\%, as shown in Table~\ref{tab:CommitVelocitydifferencebetweenGitHubandGitLab}.

\begin{table}[H]
\centering
\renewcommand{\arraystretch}{1.3}
\caption{Commit Velocity difference between GitHub and GitLab}
\label{tab:CommitVelocitydifferencebetweenGitHubandGitLab}
\begin{tabular}{|c|c|c|}
\hline
\textbf{}  & \textbf{CI/CD} & \textbf{No CI/CD}   \\ \hline
 \textbf{GitHub}  & 16.51 Hours & 27.11 Hours   \\ \hline
\textbf{GitLab}    &  22.01 & 26.7 \\ \hline
\multicolumn{3}{|c|}{Velocity Increase with CI/CD: 141.19 \%} \\ \hline
\end{tabular}
\end{table}


\subsection{\textbf{RQ2: Does integrating CI/CD impact the number of issues reported for the project?}}

In delving deeper into RQ2, as previously discussed, the use of CI/CD has an impact on the number of issues found within the project repository. Our research and analysis revealed that when CI/CD is utilized, there are generally more issues created within the repository. Although CI/CD improves and accelerates commit velocity, it is interesting to note that the number of issues within the repository would increase. However, this is not surprising as the pipeline may naturally create errors as it executes. As the project is continually deployed, the likelihood of issues increases with more commits made. The swiftness of these commits may contribute to more issues found within the repository. On average, as shown in Table~\ref{tab:IssuesdifferencebetweenGitHubandGitLab} implementing CI/CD results in a significant increase in the number of issues found within the code repository by 321.21\%.

\begin{table}[H]
\centering
\renewcommand{\arraystretch}{1.3}
\caption{Difference in the Number of Issues Between GitHub and GitLab}
\label{tab:IssuesdifferencebetweenGitHubandGitLab}
\begin{tabular}{|c|c|c|}
\hline
\textbf{}  & \textbf{CI/CD} & \textbf{No CI/CD}   \\ \hline
 \textbf{GitHub Avg Issues}  & 135.38 & 57.33   \\ \hline
\textbf{GitLab Avg Issues}    &  52.04 & 17.68 \\ \hline
\multicolumn{3}{|c|}{Issues Increase with CI/CD: 321.21 \%} \\ \hline
\end{tabular}
\end{table}


\subsection{\textbf{RQ3: Are there any significant variations in CI/CD issues between GitLab and GitHub?}}

This research question evaluates the usefulness of GitLab compared to GitHub based on the findings of this study. Our research indicates that GitHub provides more benefits than GitLab regarding commit velocity and issues found within the repository. In addition, the analysis reveals that GitHub has a higher commit velocity compared to GitLab.

When considering this question, it is also noteworthy to mention the difference in data collection between GitLab and GitHub. GitLab provides fewer tools for analyzing data within the repository compared to GitHub, which offers a more extensive suite of tools to facilitate data collection and analysis.

Finally, it is worth noting that on GitHub, a higher number of stars on a project indicates a higher likelihood of the project implementing CI/CD pipelines within the repository. However, this trend was not as evident with GitLab. The number of projects using CI/CD versus those not using CI/CD did not show a significant difference based on the number of stars on the project.

\section{Threats to Validity}

During this research project, there were several potential threats to the validity of the study. Firstly, the time taken to collect repositories resulted in a small margin being collected. Although we collected more than 12,000 repositories, this is a small number when compared to the millions of users who use these open-source repository services daily.

The main limiting factor of this project was the amount of analysis done on the data. Although the analysis presented in this paper is sufficient to yield clear and accurate outcomes, it would have been beneficial to have more time to conduct additional analysis and delve deeper into the data set. Unfortunately, additional time was not available to conduct further experiments, which could have enhanced the implications and reach of this research.

\section{Conclusion}

Many developers consider implementing CI/CD within their project pipeline because it is widely discussed. According to this study, implementing CI/CD has increased commit velocity. However, using CI/CD also significantly increases the number of issues that arise within a project. Therefore, developers must balance the benefits of faster commit velocity against the potential downsides of introducing more issues into the code base when implementing CI/CD. In the future, we would like to expand our dataset with more repositories in both sources and provide more in-depth generalized insights into our research questions.

\end{document}